\documentclass[preprint,showpacs,preprintnumbers,amsmath,amssymb]{revtex4}

\usepackage{graphicx}% Include figure files
\usepackage{dcolumn}% Align table columns on decimal point
\usepackage{bm}% bold math
\usepackage{psfig}

\def\be{\begin{equation}}
\def\ee{\end{equation}}
\def\bea{\begin{eqnarray}}
\def\eea{\end{eqnarray}}
\def\pt{\partial}

\def\al{\alpha}

\def\Gm{\Gamma}
\def\Th{\Theta}
\def\eps{\varepsilon}
\def\Dt{\Delta}

\def\om{\omega}

\def\const{\mbox{const}}

\def\mod{\mbox{mod}}
\def\ln{\mbox{ln}}

\def\dd{\mbox{d}}

\def\calE{{\cal E}}

\begin{document}

\title{Directed transport in a spatially periodic potential under periodic non-biased forcing }

\vskip 15mm

\author{Xavier Leoncini$^{1}$, Anatoly Neishtadt$^{2,3}$,
and Alexei Vasiliev$^{2}$}
  \email{valex@iki.rssi.ru}

\affiliation{$^{1}$ Centre de Physique Th\'eorique, Aix-Marseille Universit\'e,
Luminy-case 907, F-13288 Marseille Cedex 09, France, $^{2}$ Space Research Institute,
Profsoyuznaya 84/32, Moscow 117997, Russia, $^{3}$  Department of Mathematical Sciences,
Loughborough University, Loughborough, LE11 3TU, UK.}

%\date{}

\begin{abstract}
Transport of a particle in a spatially periodic harmonic potential under the influence
of a slowly time-dependent unbiased periodic external force is studied. The equations
of motion are the same as in the problem of a slowly forced nonlinear pendulum. Using
methods of the adiabatic perturbation theory we show that for a periodic external
force of general kind the system demonstrates directed (ratchet) transport in the
chaotic domain on very long time intervals and obtain a formula for the average
velocity of this transport. Two cases are studied: the case of the external force of
small amplitude, and the case of the external force with amplitude of order one. The
obtained formulas can also be used in case of a non-harmonic periodic potential.

\end{abstract}


\maketitle

\section{Introduction}

In recent years, studies of transport phenomena in nonlinear systems have been attracting
a growing interest. In particular, a large and constantly growing number of papers are
devoted to dynamics in systems which allow for directed [on average] motion under
unbiased external forces and are referred to as ratchet systems. Their intensive study
was motivated by problems of motion of Brownian particles in spatially periodic
potentials, unidirectional transport of molecular motors in biological systems, and
recognition of ``ratchet effects'' in quantum physics (see review \cite{R2002} and
references therein). Generally speaking, ratchet phenomena occur due to lack of symmetry
in the spatially periodic potential and/or the external forcing. It is interesting,
however, to understand microscopic mechanisms leading to these phenomena. A possible
approach is to neglect dissipation and noise terms and arrive at a Hamiltonian system
with deterministic forcing. Thus, one can make use of results obtained and methods
developed in the theory of Hamiltonian chaos. Many papers studying chaotic transport in
such Hamiltonian ratchets appeared in the last years (see, e.g.
\cite{JKH,Mateos,YFR,FYZ,DF,DFOYZ,HSH}). In particular, in \cite{DF} the ratchet current
is estimated in the case when there are stability islands in the chaotic domain in the
phase space of the system. The borders of such islands are ``sticky'' \cite{Zasl02} and
this stickiness together with desymmetrization of the islands are responsible for
occurrence of the ratchet transport.

We consider the problem of motion of a particle in a periodic harmonic potential
$U(q)=\om_0^2 \cos q$, where $q$ is the coordinate and $\om_0 = \const$, under the
influence of unbiased time periodic external forcing. The equations are the same as in
the paradigmatic model of a nonlinear pendulum under the action of external torque with
zero time average. We study the case when the external forcing is a time-periodic
function of large period of order $\eps^{-1}$, $0<\eps \ll 1$, and use results and
methods of the adiabatic perturbation theory. If $\eps$ is small enough, there are no
stability islands in the domain of chaotic dynamics (see \cite{NV07}). Thus, the
mechanism of ratchet transport in this system differs from one suggested in \cite{DF}.

The main objective of this paper is to find a formula for the average velocity
$V_q=\langle \dot q \rangle$ of a particle in the chaotic domain on very large time
intervals. Chaos in the system is a result of multiple separatrix crossings due to slow
variation of the external forcing. At each crossing the adiabatic invariant (``action''
of the system) undergoes a quasi-random jump (see \cite{CET,N86}). We show that these
jumps result in effective mixing and uniform distribution of the action along a
trajectory in the chaotic domain. On the other hand, direction and value of velocity
depends on the immediate value of the action. Thus, to find the average velocity of
transport on time intervals of order or larger than the mixing time, we find formulas for
displacement in $q$ at a given value of the action and then integrate them over the
interval of values of the action corresponding to the chaotic domain. We demonstrate that
for an external force of general kind (i.e. with zero time average but lowered time
symmetry, cf. \cite{FYZ}), there is directed transport in the system and obtain an
analytic formula for the average velocity $V_q$ of this transport.

In Section 2, we obtain the main equations in the case of external force of small
amplitude and describe the diffusion of adiabatic invariant due to multiple separatrix
crossings. This diffusion makes dynamics chaotic in a large domain of width of order 1.
In Section 3, we derive the formula for the average velocity of transport on very long
time intervals (of order of typical diffusion time and larger) and check it numerically.
In Section 4, we consider the case when the external forcing is not small, of order 1.
Width of the chaotic domain in this situation is large, of order $\eps^{-1}$. In this
case chaos arises as a result of scattering of the adiabatic invariant on the resonance.
Formula for the average velocity of transport in this case turns out to be much simpler
than in the case of external forcing of small amplitude.

\section{Main equations. Diffusion of the adiabatic invariant}

\begin{figure}[htbp]
%\vspace*{-3mm}
\psfig{file=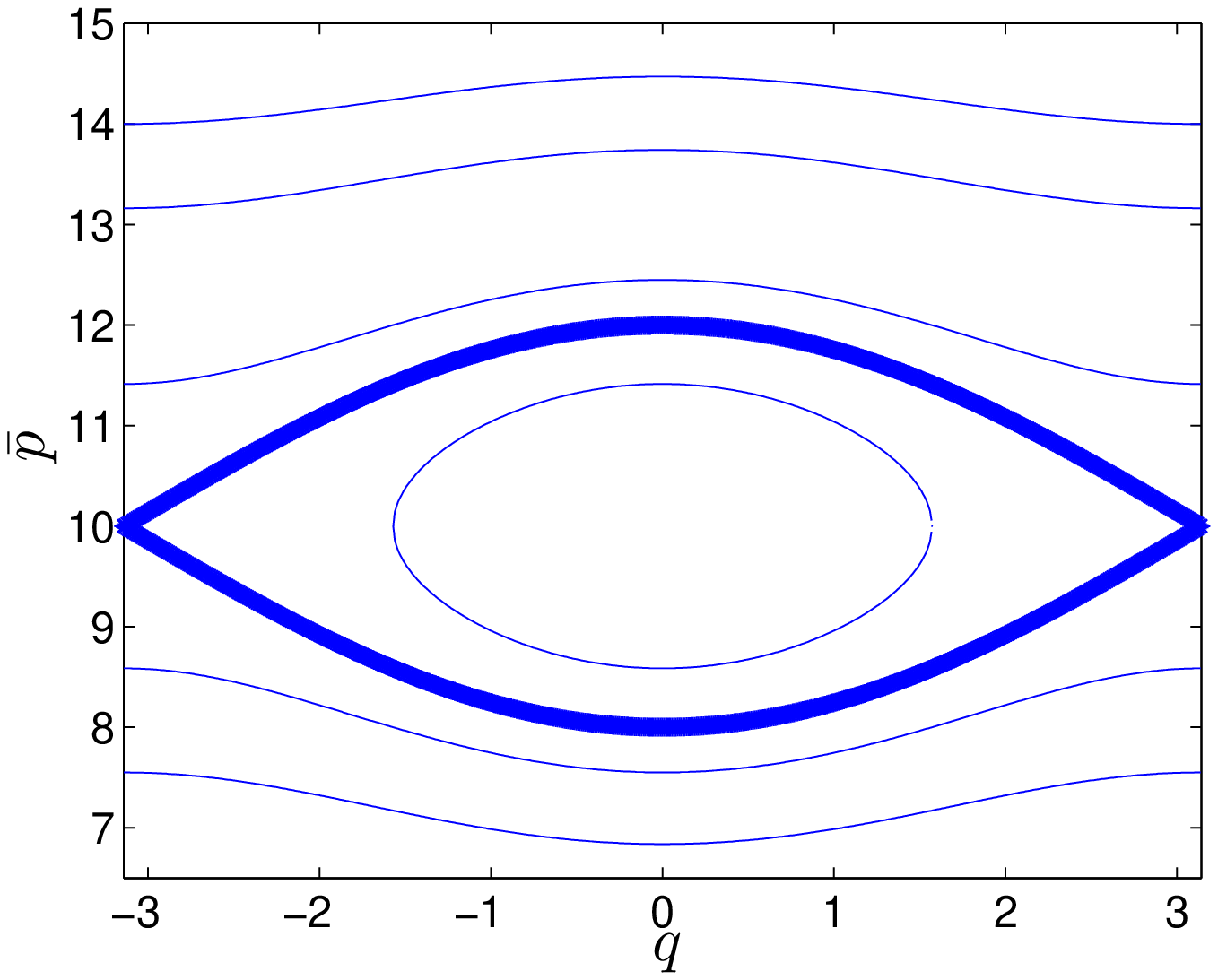,width=200pt} \caption{\small{Phase trajectories of the system
(\ref{1.3}) with sample function $F(\tau) = A\left(1+2\exp\left[-\al(\sin \tau)^2\right]
\right)$, $A=10,\, \al=8,\, \tau=\pi/2,\, \om_0=1$. The bold line is the separatrix. }}
\label{portrait}
\end{figure}

We start with a basic equation describing a nonlinear pendulum under the influence of an
external force $\tilde f(t)$:
\be
\ddot{q} +\om_0^2 \sin q = \tilde f(t).
\label{1.1}
\ee
We assume that $\tilde f(t)$ is small in amplitude periodic function of time of large
period: $\tilde f(t)\equiv \eps f(\tau)= \eps f(\tau + 2\pi)$, where $\tau = \eps t$.
Moreover, we assume that $\tilde f(t)$ is a function with zero time average. This system
is Hamiltonian, with time-dependent Hamiltonian function
\be
H = \frac{p^2}{2} - \om_0^2 \cos q - \eps f(\eps t) q.
\label{1.2}
\ee
We make a canonical transformation of variables $(p,q)\mapsto(\bar p, \bar q)$ using
generating function $W=(\bar p - F(\eps t))q$, where $F(\tau) = - \int f(\tau) \dd \tau$.
Thus, $F(\tau)$ is a periodic function defined up to an additive constant, which we are
free to choose. To make the following presentation more clear, we choose this constant in
such a way that the minimal value of $F$ is $F_{min} > 4\om_0/\pi$. Note that $\bar q
\equiv q$. After this transformation, Hamiltonian of the system acquires the form (bars
over $q$ are omitted):
\be
H = \frac{(\bar p - F(\tau))^2}{2} - \om_0^2 \cos q.
\label{1.3}
\ee
Phase portrait of the system at a frozen value of $\tau$ (we call it the unperturbed
system) is shown in Fig.\ \ref{portrait}. The separatrix divides the phase space into the
domains of direct rotations (above the upper branch of the separatrix), oscillations
(between the separatrix branches), and reverse rotations (below the lower branch of the
separatrix). Introduce the ``action'' $I$ associated with a phase trajectory of the
unperturbed system on this portrait. In the domains of rotation, $I$ equals an area
between the trajectory, the lines $q=-\pi,\, q=\pi$, and the axis $\bar p=0$, divided by
$2\pi$; in the domain of oscillations, this is an area surrounded by the trajectory
divided by $2\pi$. It is known that $I$ is an adiabatic invariant of (\ref{1.3}): far
from the separatrix its value is preserved along a phase trajectory with the accuracy of
order $\eps$ on long time intervals (see, e.g., \cite{AKN}).

Location of the separatrix on the $(q,\bar p)$-plane depends on the value of $F(\tau)$.
As $\tau$ slowly varies, the separatrix slowly moves up and down, and phase points cross
the separatrix and switch its regime of motion from direct rotations to reverse rotations
and vice versa. Recall known results on variation of the adiabatic invariant when a phase
point crosses the separatrix. The area surrounded by the separatrix is constant, and
hence, capture into the domain of oscillations is impossible in the first approximation
(in the exact system, only a small measure of initial conditions correspond to phase
trajectories that spend significant time in this domain; thus their influence on the
transport is small). To be definite, consider the situation when the separatrix on the
phase portrait slowly moves down. Thus, phase points cross the separatrix and change the
regime of motion from reverse rotation to direct rotation. Let the action before the
separatrix crossing at a distance of order 1 from the separatrix be $I = I_-$ and let the
action after the crossing (also at a distance of order 1 from the separatrix) be $I=I_+$.
In the first approximation, we have $I_+ = I_- + 8\om_0/\pi$, i.e. the action increases
by the value of the area enclosed by the separatrix divided by $2\pi$ (see, e.g.,
\cite{Ch59,N75}). We shall call this change in the action a ``geometric jump''. If the
separatrix contour slowly moves up, and a phase point goes from the regime of direct
rotation to the regime of reverse rotation, the corresponding value of the action
decreases by the same value $8\om_0/\pi$. Thus, in this approximation, the picture of
motion looks as follows. While a phase point is in the domain of reverse rotation, the
value of $I$ along its trajectory stays constant: $I=I_-$. After transition to the domain
of direct rotation, this value changes by the value of the geometric jump. The transition
itself in this approximation occurs instantaneously. After the next separatrix crossing,
the adiabatic invariant changes again by the value of the geometric jump, with the
opposite sign, and returns to its initial value $I_-$. We call this approximation
adiabatic.

In the next approximation, the value of action at the separatrix crossing undergoes a
small additional jump. Consider for definiteness the case when the separatrix contour on
the phase portrait moves down, and $I_-$ and $I_+$ are measured when it is in its
uppermost and lowermost positions, accordingly. Results of \cite{CET,N86} imply the
following formula for the jump in the adiabatic invariant:
\bea
2\pi(I_+ - I_-) &=& 16\om_0 + 2a(1-\xi)\eps\Th \ln(\eps\Th)
\nonumber \\
&+& a\eps\Th \ln\frac{2\pi (1-\xi)}{\Gm^2(\xi)} -2b\eps\Th(1-\xi),
\nonumber
\eea
where $a=\om_0^{-1},\; b=\om_0^{-1}\ln(32\om_0^2),\; \Th = 2\pi F^{\prime}(\tau_*)$. Here
$F^{\prime}$ is the $\tau$-derivative of $F$, $\tau_*$ is the value of $\tau$ at the
separatrix crossing found in the adiabatic approximation, $\Gm(\cdot)$ is the
gamma-function. Value $\xi$ is a so-called pseudo-phase of the separatrix crossing; it
strongly depends on the initial conditions and can be considered as a random variable
uniformly distributed on interval $(0,1)$ (see, e.g., \cite{N86}). Thus, value of the
jump in the adiabatic invariant at the separatrix crossings has a quasi-random component
of order $\eps\ln\eps$.

\begin{figure}[ht]
%\vspace*{-3mm}
\begin{tabular}{cc}
\hspace{5mm}\psfig{file=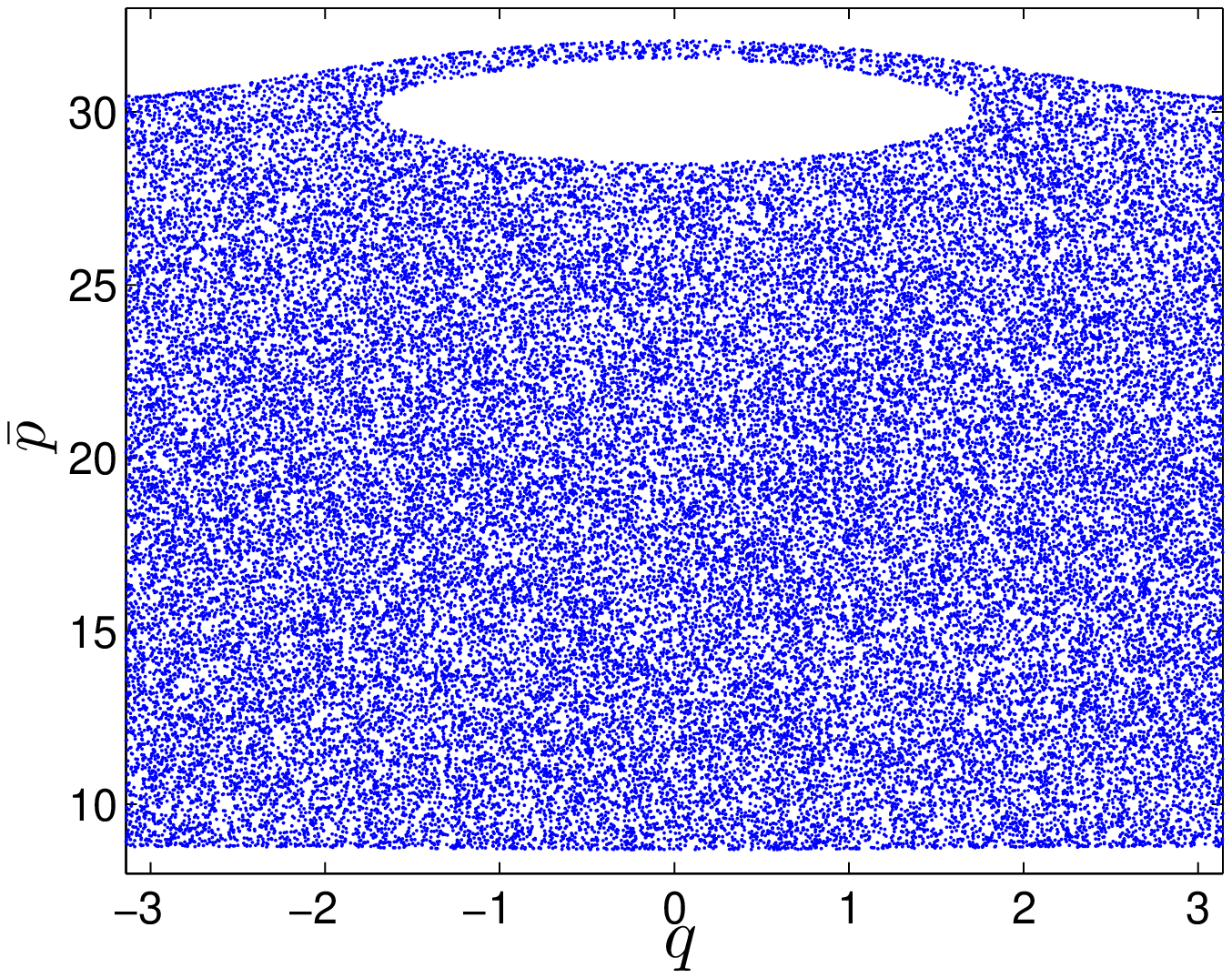,width=200pt} &
\hspace{5mm}\psfig{file=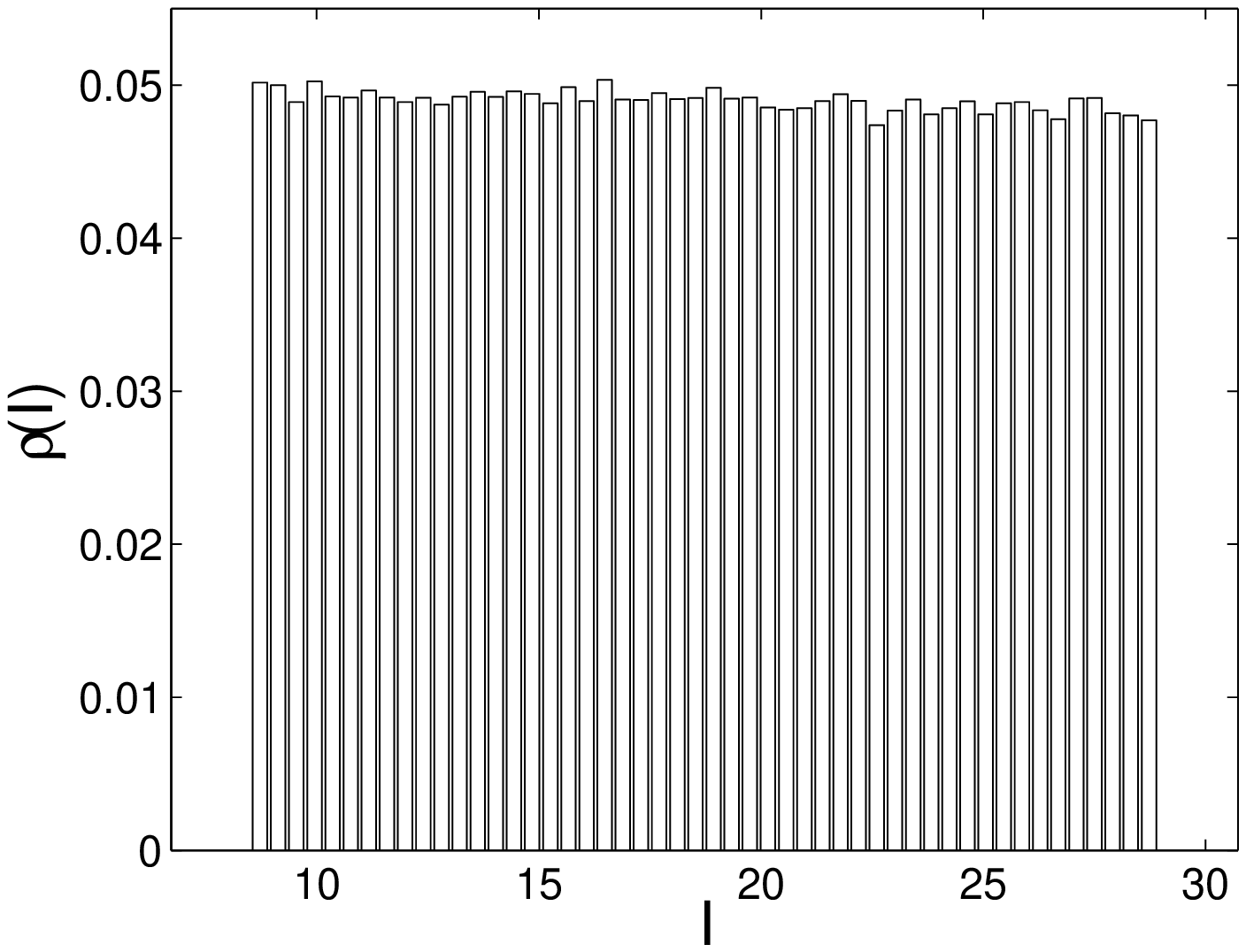,width=200pt}
\end{tabular}
\caption{\small{Left panel: Poincar\'e section at $\tau=0\, \mod\,2\pi$ of a long phase
trajectory ($5 \cdot 10^4$ dots) . All the points are mapped onto the interval
$q\in(-\pi,\pi)$. $F(\tau) = A\left(1+2\exp\left[-\al(\sin \tau)^2\right] \right)$ with
$A=10, \al=16, \eps=0.005, \om_0=1$. The empty region in the chaotic sea corresponds to
phase points eternally locked in the domain of oscillations; they never enter the chaotic
domain and do not participate in the transport. Right panel: Histogram of $I$ on the
segment $(I_{min},I_{max} - 8\om_0/\pi))$ along the same phase trajectory. }}
\label{density}
\end{figure}

Accumulation of small quasi-random jumps due to multiple separatrix crossings produces
diffusion of adiabatic invariant (see, e.g., \cite{N86}). On a period of $F(\tau)$ (after
two separatrix crossings) the action changes by a value of order $\eps\ln\eps$. Hence,
after $N\sim \eps^{-2}(\ln\eps)^{-2}$ separatrix crossings the adiabatic invariant varies
by a value of order one. As a result, in time of order $t_{diff}\sim
\eps^{-3}(\ln\eps)^{-2}$ the value of adiabatic invariant is distributed in all the range
of values corresponding to the domain where phase points cross the separatrix on the
phase plane; its distribution is close to the uniform one. We have checked this fact
numerically for a sample function $F(\tau)$ at various parameter values. Poincar\'e
sections and distribution histograms of $I$ in all the cases look similar; see an example
in Fig. \ref{density}.

\section{Average velocity of the transport}

Our aim is to find a formula for average velocity $V_q$ along a phase trajectory on time
intervals of order $t_{diff}$ or larger. We first only take into consideration the
geometric jumps, and afterwards, to obtain the final result, we take into account the
mixing due to small quasi-random jumps. To simplify the consideration, assume that
function $F(\tau)$ has one local minimum $F_{min}$ and one local maximum $F_{max}$ on the
interval $(0,2\pi)$. The main results are valid without this assumption.

Introduce $\tilde I$, defined in the domains of rotation as follows: it equals the area
bordered by the trajectory, the line $\bar p=F(\tau)$, and the lines $q=-\pi,\, q=\pi$,
divided by $2\pi$. Thus, $\tilde I=|F(\tau) - I|$. Frequency of motion in the domains of
rotation is $\om(\tilde I)$, where $\om(\tilde I)$ at $\tilde I> 4\om_0/\pi$ is the
frequency of rotation of a standard nonlinear pendulum with Hamiltonian $H_0 = p^2/2 -
\om_0^2 \cos q$, expressed in terms of its action variable $\tilde I$. We do not need an
explicit expression for function $\om(\tilde I)$. From Hamiltonian (\ref{1.3}) we find
$\dot q = \bar p - F(\tau)$. Consider a phase trajectory of the system frozen at
$\tau=\bar \tau$ in a domain of rotation. Let the value of action on this trajectory be
$I=I_0$. Then the value of $\dot q$ averaged over a period $T$ of rotation equals
$\int^T_0 |\dot q| \, \dd t/T = 2\pi/T = \om(|F(\bar\tau) - I_0|)$.

Now consider a long phase trajectory in the case of slowly varying $\tau$. Let on the
interval $(\tau_1,\tau_2)$ a phase point of (\ref{1.3}) be below the separatrix contour.
In the adiabatic approximation, the value $I_0$ of the adiabatic invariant along its
trajectory is preserved on this interval. Hence, at $\tau \in (\tau_1,\tau_2)$ we have
\be
2\pi F(\tau) - 2\pi I_0 \geq 8\om_0,
\label{2.2}
\ee
and the equality here takes place at $\tau = \tau_1$ and $\tau = \tau_2$. In the process
of motion on this time interval, $q$ changes (in the main approximation) by a value
\be
\Dt q_-(I_0) = - \frac{1}{\eps}\,\int_{\tau_1}^{\tau_2} \om(F(\tau) - I_0) \dd \tau.
\label{2.3}
\ee

On the interval $(\tau_2,\tau_1+2\pi)$ the phase trajectory is above the separatrix
contour, and the value of the adiabatic invariant equals $\hat I_0 = I_0 + 8\om_0/\pi$
due to the geometric jump. On this interval we have
\be
2\pi F(\tau) - 2\pi I_0 \leq 8\om_0.
\label{2.4}
\ee
In the process of motion on this time interval, $q$ changes by a value
\be
\Dt q_+(I_0) = \frac{1}{\eps}\,\int_{\tau_2}^{\tau_1+2\pi} \om(|F(\tau) - \hat I_0|) \dd
\tau .
\label{2.5}
\ee
Total displacement in $q$ on the interval $(\tau_1,\tau_1+2\pi)$ equals $\Dt q (I_0) =
\Dt q_-(I_0) + \Dt q_+(I_0)$, and the average velocity on this interval is $\eps\Dt
q(I_0)/(2\pi)$.

Consider now the motion on a long enough time period $\Dt t \sim t_{diff}$. Due to the
diffusion in the adiabatic invariant described above, in this time period values of
$I_0$, defined as a value of $I$ when the phase point is {\it below} the separatrix
contour, cover the interval $(I_{min},I_{max} - 8\om_0/\pi)$. Here $I_{min} = F_{min} -
4\om_0/\pi$ and $I_{max} = F_{max} + 4\om_0/\pi$.  Assuming that the distribution of $I$
on this interval is uniform, to find the average velocity, we integrate $\eps\Dt
q(I_0)/(2\pi)$ over this interval. Integrating (\ref{2.3}) over $I_0$ and changing the
order of integration we find
\bea
\int_{I_{min}}^{I_{max}-8\om_0/\pi}\!\! \!\!\! \Dt q_- \dd I_0 \!\!\!&=&\!\!\!
-\frac{1}{\eps}\!\!\int_0^{2\pi}\!\!\!\! \dd \tau
\int_{I_{min}}^{F(\tau)-4\om_0/\pi}\!\!\!\!\!\!\!\! \om(F(\tau) - I_0) \dd I_0
\nonumber \\
&=& -\frac{1}{\eps}\int_0^{2\pi} \dd \tau \int_{4\om_0/\pi}^{F(\tau) - I_{min}}
\om(\eta)\dd \eta.
\nonumber
\eea
Now we take into account the equality $\om(\tilde I) = \pt H_0(\tilde I)/\pt \tilde I$
(recall that $H_0(\tilde I)$ is the Hamiltonian of a nonlinear pendulum as a function of
its action variable) and obtain
\be
-\!\!\int_0^{2\pi} \!\!\! \dd \tau \int_{4\om_0/\pi}^{F(\tau) \!-\! I_{min}}\!\!\!\!\!
\om(\eta)\dd \eta \!=\! -\!\int_0^{2\pi}\!\!\!\! \left(H_0(F(\tau)\!-\!I_{min})\! -\!
H_0^s \right) \dd \tau,
\label{2.7}
\ee
where $H_0^s$ is the value of $H_0$ on the separatrix. Similarly, integrating (\ref{2.5})
we obtain
\be
\int_{I_{min}}^{I_{max}-8\om_0/\pi}\!\!\!\! \Dt q_+ \dd I_0 = \frac{1}{\eps}\int_0^{2\pi}
\!\!\! \left(H_0(I_{max} \!-\! F(\tau)) \!-\! H_0^s \right) \dd \tau.
\label{2.8}
\ee
Adding (\ref{2.7}) to (\ref{2.8}) and dividing by $2\pi(F_{max}-F_{min})/\eps$ we find
the expression for the average velocity $V_q$ of transport on long time intervals:
\bea
V_q \!&=&\! \frac{1}{2\pi(F_{max}-F_{min})}
\nonumber \\
&\times & \!\! \int_0^{2\pi}\!\!\!\!\left(H_0(I_{max} \!-\!F(\tau))\!\! -\!\!
H_0(F(\tau)\!-\! I_{min})\right) \dd \tau.
\label{2.9}
\eea
In (\ref{2.9}), $H_0(I)$ can be found as the inverse function to $\tilde I(h)$, which
defines action as a function of energy in domains of rotation of a nonlinear pendulum.
For the latter function, the following formula holds (see, e.g., \cite{SUZ}):
\be
\tilde I(h) = \frac{4}{\pi}\om_0 \kappa \, \calE \left(1/\kappa \right), \; \kappa \geq
1,
\label{2.10}
\ee
where $\kappa^2 = (1+ h/\om_0^2)/2$, $\calE(\cdot)$ is the complete elliptic integral of
the second kind. If function $F(\tau)$ has several local extremes on the interval
$(0,2\pi)$, $F_{min}$ and $F_{max}$ in (\ref{2.9}) are the smallest and largest values of
$F$ respectively.

\begin{table}[t]
\begin{tabular}{|l|c|c|c|}
\hline & $\al=4$ & $\al=8$ & $\al=16$ \\
\hline $\eps=0.1$ & 4.721 & 6.756 & 8.363 \\
$\eps = 0.05$ & 4.446 & 6.681 & 8.076 \\
$\eps = 0.01$ & 4.298 & 6.211 & 7.442 \\
$\eps =0.005$ & 4.598 & 6.702 & 8.202 \\
\hline \hline $V_q^{theor}$ & 4.393 & 6.679 & 8.110 \\
\hline
\end{tabular}
\caption{\small Numerically found values of $V_q$ corresponding to various values of
parameters $\eps, \, \al$ (four upper rows, $A=10, \, \om_0=1$) and theoretical values
$V_q^{theor}$ obtained according to  (\ref{2.9}) (the bottom row). }
\label{table1}
\end{table}

\begin{figure}[ht]
\begin{tabular}{ccc}
\hspace*{-12mm}\psfig{file=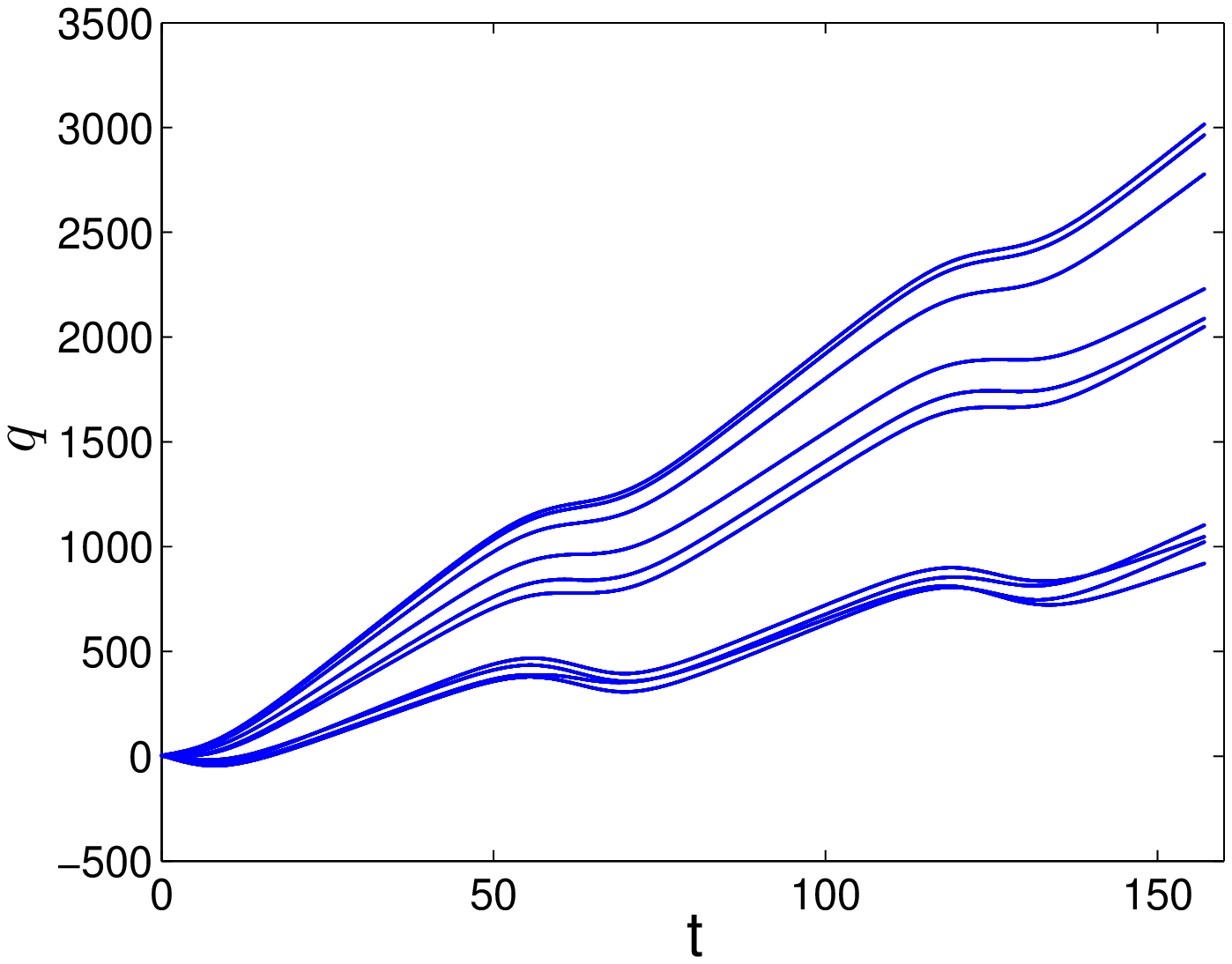,width=180pt} &
\hspace*{-5mm}\psfig{file=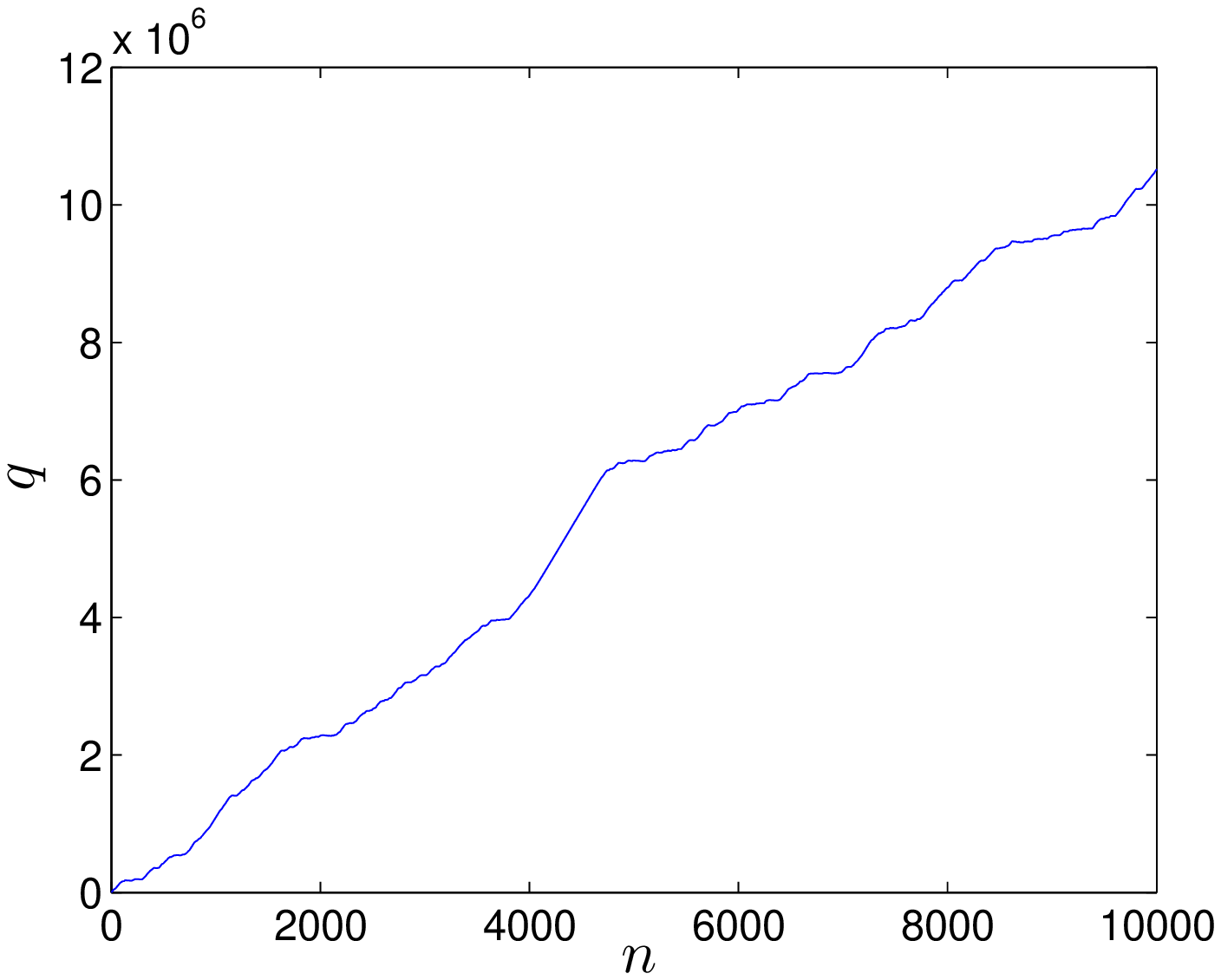,width=180pt} &
\hspace*{-5mm}\psfig{file=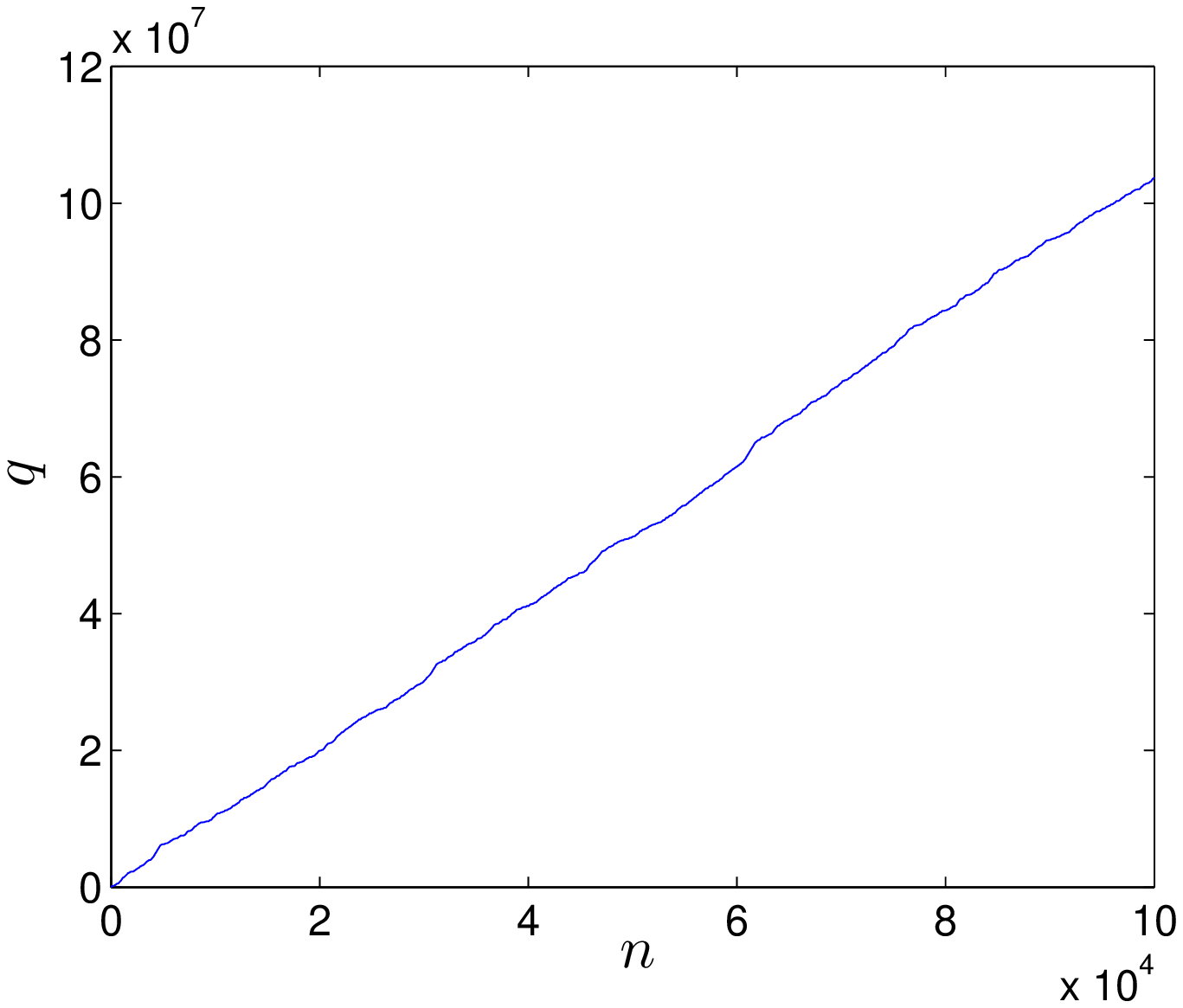,width=180pt}
\end{tabular}
\caption{Left panel: $q$ against $t$ for ten different initial conditions (comparatively
short time interval), $\alpha=4, \varepsilon=0.05$. Central panel: $q$ against the number
of periods of the external force for a sample trajectory ($10^4$ periods), $\alpha=16,
\varepsilon=0.05$. Right panel: $q$ against the number of periods of the external force
for the same trajectory ($10^5$ periods), $\alpha=16, \varepsilon=0.05$. Parameter $A=10$
in all cases. }
\label{transport}
\end{figure}

It can be seen from (\ref{2.9}) that for function $F(\tau)$ of general type $V_q$ is not
zero, and hence there is the directed transport in the system. We checked this formula
numerically for a sample function $ F(\tau) = A\left(1+2\exp\left[-\al(\sin
\tau)^2\right] \right) ,\, \al>0$ at various values of parameters $\eps$ and $\al$.
Typical plots of $q$ against time $t$ are shown in Fig. \ref{transport}.

The results of numerical checks of formula (\ref{2.9}) are represented in Table
\ref{table1}. To find numerical values of $V_q$ presented in the table, we integrated the
system with Hamiltonian (\ref{1.3}) on a long time interval $\Dt t = 2\pi \cdot
10^6/\eps$ with a constant time step of $\pi/100$ (5th order symplectic scheme
\cite{MA}). Use of a symplectic scheme for long time simulations of Hamiltonian systems
is necessary in order to ensure that creeping numerical error do not end up washing off
the invariant tori bounding the chaotic domain. The table demonstrates satisfactory
agreement between the formula and the numerics.

Finally, we note that formula (\ref{2.9}) can be used also in the case of arbitrary
(non-harmonic)  spatially-periodic time independent potential in place of the term $-
\om_0^2 \cos q$ in (\ref{1.2}), (\ref{1.3}). Of course, in this case function $H_0$ is
different from the Hamiltonian of the nonlinear pendulum, but it always can be found,
at least numerically.

\section{The case of not small external forcing}

In this section we study the case when the external forcing is not small. In this case,
amplitude of function $\tilde f(t)$ in (\ref{1.1}) is a value of order one: $\tilde
f(t)\equiv f(\tau)= f(\tau + 2\pi)$. The equations of motion are:
\be
\dot q = p, \,\,\,\dot p=-\om_0^2 \sin q +f(\tau), \,\,\, \dot \tau = \eps.
\label{3.1}
\ee
One can see from the second equation, that magnitude of momentum $p$ can reach values of
order $\eps^{-1}$. Making the canonical transformation with generating function $W=(\bar
p - \eps^{-1}F(\eps t))q$, where $F(\tau)$ is defined in Section II, we obtain the
Hamiltonian:
\be
H = \frac{(\bar p - \eps^{-1}F(\tau))^2}{2} - \om_0^2 \cos q.
\label{3.2}
\ee
Introduce $\tilde p=\eps \bar p$ and rescaled time $\tilde t=\eps^{-1} t$. We denote the
derivative w.r.t. $\tilde t$ with prime and thus obtain:
\be
q^{\prime} = \tilde p -F(\tau), \,\,\,\tilde p^{\prime}=-\eps^2\om_0^2\sin q , \,\,\,
\tau^{\prime} = \eps^2.
\label{3.3}
\ee
This is a system in a typical form for application of the averaging method. We average
over fast variable $q$ and obtain the averaged system
\be
\tilde p^{\prime}=0, \,\,\, \tau^{\prime} = \eps^2.
\label{3.4}
\ee
The averaged system describes the dynamics adequately everywhere in the phase space
except for a small neighborhood of the resonance at $\tilde p -F(\tau) =0$, where the
``fast'' variable $q$ is not fast. When a phase trajectory of the averaged system crosses
the resonance, value of the adiabatic invariant (in this case it coincides with $\tilde
p$) undergoes a quasi-random jump of typical order $\sqrt{\eps^2}=\eps$ (see, e.g.,
\cite{N97,AKN}). Thus, the situation in this case is similar to one studied in Section 2,
but the typical value of a jump is of order $\eps$, and, accordingly, the typical
diffusion time is $t_{diff} \sim \eps^{-3}$. Phase trajectories of the averaged system
that cross the resonance correspond to values of $\tilde p$ belonging to the interval
$(F_{min},F_{max})$. Therefore the chaotic domain of the exact system is, in the main
approximation, a strip $F_{min}\le \tilde p \le F_{max}$. Captures into the resonance
followed by escapes from the resonance (see \cite{N97,AKN}) are also possible in this
system. However, probability of capture is small, of order $\eps$, and hence impact of
these phenomena on the transport is small.

\begin{figure}[ht]
%\vspace*{-3mm}
\begin{tabular}{cc}
\hspace{5mm}\psfig{file=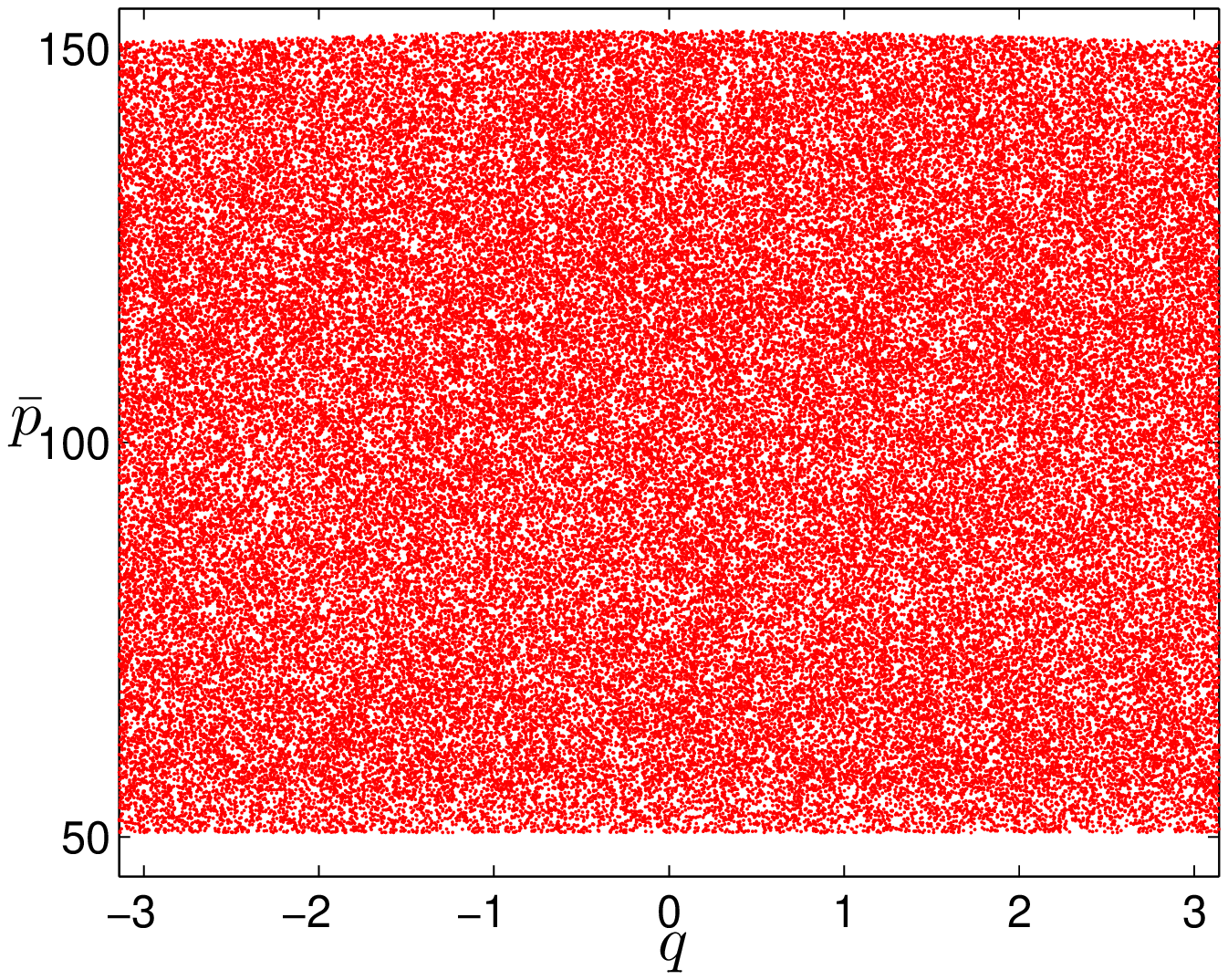,width=200pt} &
\hspace{5mm}\psfig{file=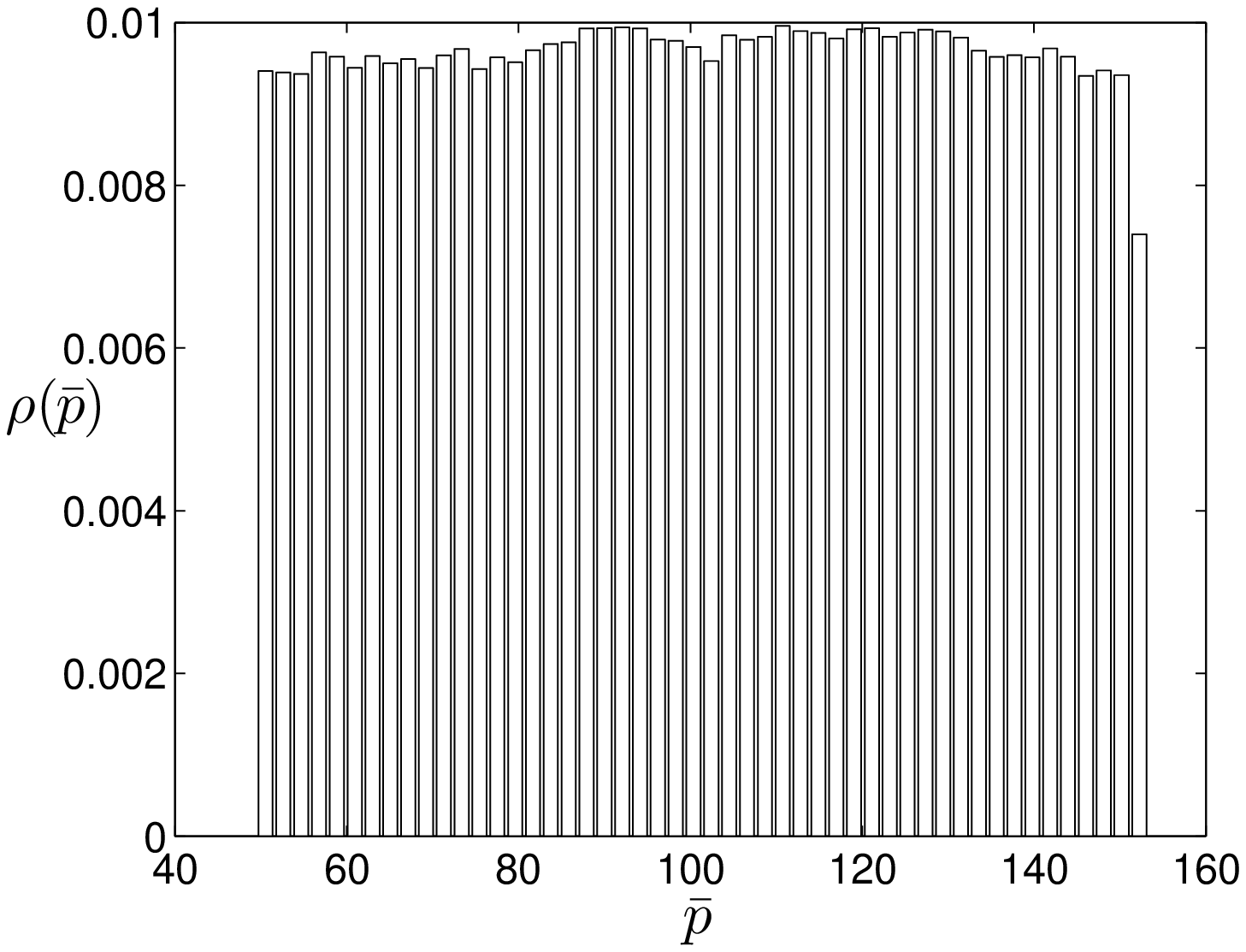,width=200pt}
\end{tabular}
\caption{\small{Left panel: Poincar\'e section at $\tau=0\, \mod\,2\pi$ of a long phase
trajectory ($5 \cdot 10^4$ dots) of system (\ref{3.2}). All the points are mapped onto
the interval $q\in(-\pi,\pi)$. $F(\tau) = A\left(1+2\exp\left[-\al(\sin \tau)^2\right]
\right)$ with $A=0.5, \al=4, \eps=0.01, \om_0=1$. Right panel: Histogram of $\bar p$
along the same phase trajectory. }}
\label{density2}
\end{figure}

In Fig. \ref{density2} we represent a sample of Poincar\'e section of a long phase
trajectory of (\ref{3.2}) and the corresponding histogram of $\bar p$ for this
trajectory. The plots show that  the distribution of $\bar p$ is close to the uniform
one.

Like in Section 3, one can find value $\Dt q$ on one period of perturbation, then average
over the range of adiabatic invariant corresponding to the chaotic domain, and find the
average velocity of transport in this case. Thus we find:
\be
\Dt q = \int_0^{2\pi/\eps} p \dd t = \int_0^{2\pi/\eps} \left(\bar p -
\frac{F(\tau)}{\eps}\right) \dd t = \frac{1}{\eps^2} \int_0^{2\pi} (\tilde p - F(\tau))
\dd \tau.
\label{3.5}
\ee
To find $V_q$, we have to integrate this expression over $\tilde p$ from $F_{min}$ to
$F_{max}$ (i.e., over the chaotic domain) and divide the result by $(F_{max}-F_{min})$
and by the length of the period of the external forcing $2\pi/\eps$. Thus we obtain
\be
V_q = \frac{\eps}{2\pi (F_{max} - F_{min})} \int_{F_{min}}^{F_{max}} \Dt q \, \dd \tilde
p.
\label{3.6}
\ee
Substituting $\Dt q$ from (\ref{3.5}) and integrating, one straightforwardly obtains
\be
V_q = \frac{1}{4\pi\eps} \, \int_0^{2\pi} \left( F_{max} + F_{min} -2 F(\tau)\right) \dd
\tau.
\label{3.7}
\ee
Another possibility to find $V_q$ is to use already obtained formula (\ref{2.9}). This
way leads to the same result. When using (\ref{2.9}), one should keep in mind that in the
considered case $H_0$ in this formula is the Hamiltonian of the averaged system, i.e.
$H_0=I^2/2$, $I_{max} = F_{max}, \,\, I_{min} = F_{min}$, and that according to
(\ref{3.2}) we should put $\eps^{-1} F$ instead of $F$ everywhere in the formula. Thus,
(\ref{2.9}) is much simplified, and we again arrive at formula (\ref{3.7}). Factor
$\eps^{-1}$ is due to the fact that a phase point spends significant time moving at large
velocities corresponding to $p \sim \eps^{-1}$.

Note that formula (\ref{3.7}) can be rewritten in a more elegant form as:
\be
V_q = \frac{1}{\eps}\left( \frac{F_{max} + F_{min}}{2} - \langle F(\tau)\rangle \right),
\label{3.8}
\ee
where the angle brackets denote time average. The results of numerical checks of the
formula are represented in Table \ref{table2}. Like in the previous Section, we
integrated the system with Hamiltonian (\ref{3.2}) on a long time interval $\Dt t = 2\pi
\cdot 10^6/\eps$.

Remarkably, formula (\ref{3.8}) is valid for arbitrary smooth periodic potential (not
necessarily harmonic) in place of the term $- \om_0^2 \cos q$ in (\ref{1.2}),
(\ref{1.3}). The  potential may also depend periodically on time with the same period
as that of the external force.

\begin{table}[t]
\begin{tabular}{|l|c|c|c|}
\hline & $\al=1$ & $\al=2$ & $\al=4$ \\
\hline $\eps=0.1$ & 0.046 & 0.128 & 0.253 \\
$\eps = 0.05$ & 0.046 & 0.112 & 0.225 \\
$\eps = 0.01$ & 0.0353 & 0.1050 & 0.2044 \\
$\eps =0.005$ & 0.0369 & 0.1081 & 0.1916 \\
\hline \hline $\eps V_q^{theor}$ & 0.0389 & 0.1018 & 0.2006 \\
\hline
\end{tabular}
\caption{\small Numerically found values of $\eps V_q$ corresponding to various values of
parameters $\eps, \, \al$ in system (\ref{3.2}) for $F(\tau) =
A\left(1+2\exp\left[-\al(\sin \tau)^2\right] \right)$  (four upper rows, $A=0.5, \,
\om_0=1$). In the bottom row theoretical values $\eps V_q^{theor}$ obtained according to
(\ref{3.8}) are shown. }
\label{table2}
\end{table}

\section{Summary}

To summarize, we have considered the phenomenon of the directed transport in a spatially
periodic harmonic potential adiabatically influenced by a periodic in time unbiased
external force. We have shown that for the external force of a general kind the system
exhibits directed transport on long time intervals.  Direction and average velocity of
the transport in the chaotic domain are independent of initial conditions and determined
by properties of the external force. We studied two different cases: the case of small
amplitude of the external force and the case, when this amplitude is a value of order
one. We have obtained an approximate formula for average velocity of the transport and
checked it numerically. The final formulas (\ref{2.9}) and (\ref{3.8}) are valid for any
smooth periodic potential (not necessarily harmonic one).

\section*{Acknowledgements}

The work was partially supported by the RFBR grants 06-01-00117 and NSh 691.2008.1. A.V.
thanks Centre de Physique Th\'eorique in Luminy and Ricardo Lima for hospitality in the
fall of 2007 and numerous discussions.

\end{document}